\documentstyle[eqsecnum,preprint,aps]{revtex}

\def\pmb#1{\setbox0=\hbox{$#1$}%
\kern-.025em\copy0\kern-wd0
\kern.05em\copy0\kern-\wd0
\kern-.025em\raise.0433em\box0 }

\def\be{\begin{equation}}
\def\ee{\end{equation}}
\def\half{{\textstyle\frac{1}{2}}}
\def\fpg{{\frac{1}{4 \pi G}}}

\def\square{\kern1pt\vbox{\hrule height1.2pt\hbox{\vrule width1.2pt\hskip3pt
\vbox{\vskip 6pt}\hskip3pt\vrule width0.6pt}\hrule height0.6pt}\kern1pt}

\begin{document}

\def\footnoterule{\hrule width \hsize}


\title{Quantum States of String-Inspired Lineal Gravity\footnotemark[1]}

\footnotetext[1]
{This work is supported in part
by funds provided by the U.S.~Department of Energy (D.O.E.)
under contract \#DE-AC02-76ER03069 (RJ),
by the U.S.\ National Science Foundation (N.S.F.) under contract
\#PHY-89-15286
and by the Swiss National Science Foundation (DC).} 

\author{D. Cangemi}
\medskip
\address{Physics Department, University of California at Los Angeles\\
405 Hilgard Ave., Los Angeles, CA ~90024--1547}

\bigskip

\author{R. Jackiw}
\medskip

\address{Center for Theoretical Physics, Laboratory for Nuclear Science
and Department of Physics \\
Massachusetts Institute of Technology, Cambridge, MA ~02139--4307}


\maketitle

\begin{abstract}
We construct quantum states for a (1+1) dimensional gravity-matter model that
is also a gauge theory based on the centrally extended Poincar\'e group.
Explicit formulas are found, which exhibit interesting structures.  For
example wave functionals are gauge invariant except for a gauge non-invariant
phase factor that is the Kirillov-Kostant 1-form on the (co-) adjoint orbit of
the group.  However no evidence for gravity-matter forces is found.
\end{abstract}

\setcounter{page}{0}
\thispagestyle{empty}

\vfill
\centerline{Submitted to: {\em Physical Review D\/} {\bf 15}}
\vfill
\hbox to \hsize{CTP\#2278 \hfil March 1994}
\hbox to \hsize{UCLA/94/TEP/1 \hfil}
\eject

\widetext

\section{Introduction}
\label{sec:1}

The string-inspired model for lineal gravity \cite{1,2} has been studied in
the last few years with the aim of gleaning useful information about black
hole physics.  Even though many papers have been published, the quantum
mechanical theory has not been solved; only semi-classical analyses of
uncertain validity have been carried out.  Of course the obstacle to a
complete quantal solution is the intractability of quantum gravity, which
persists even when the world has been dimensionally reduced to one, lineal
dimension.

In this paper we report new results in our approach to the problem of
quantizing string-inspired lineal gravity, once it has been reformulated as a
gauge theory \cite{2} of the extended Poincar\'e group \cite{3,4}.
Specifically when point particles are coupled to the gravitational degrees of
freedom, the quantum states can be constructed, and we present explicit
wave functionals for the one- and many-particle cases.  The pure-gravity
wave functionals, which had been previously found \cite{5,6}, are also
discussed.

The rationale for a gauge theoretical formulation
of gravity theory
is the hope that familiar
techniques for quantizing gauge theories can be successfully employed, thereby
circumventing apparently intractable problems of quantum gravity
(diffeomorphism constraints, Wheeler-DeWitt equation, {\it etc.}).
Our success with the point particle problem encourages optimism.

However, another reason should be put forward in favor of the gauge
theoretical formulation.  When the string-inspired model was first proposed,
the gravity action was taken to be
\be
\bar{I}_G = \frac{1}{4\pi G} \int d^2 x ~ \sqrt{-\bar{g}} \, e^{-2\varphi} \,
\left( \bar{R} + 4 \bar{g}^{\mu\nu} \, \partial_\mu \, \varphi
\, \partial_\nu \,
\varphi - \lambda \right)
\label{1.1}
\ee
where $\varphi$ is the ``dilaton'' field and $\lambda$ a
cosmological constant.
(Bars are used to distinguish the above from a subsequent, redefined
expression; see below.)
Matter is coupled only to $\bar{g}_{\mu\nu}$ in a conformally invariant
manner, so the trace of the energy-momentum tensor is given solely by its
quantum anomaly, proportional to the scalar curvature $\bar{R}$, which
according to the dynamics implied by (\ref{1.1}) is a non-trivial quantity.
Since (1+1)-dimensional semi-classical Hawking radiation is governed by the
trace anomaly, the above results would indicate that black hole phenomena,
Hawking radiation, {\it etc.}\ arise in this model.  Indeed a ``black hole''
classical solution to the equations has been identified \cite{1,2}.

Subsequently,
it was also realized that a redefinition of variables
\begin{mathletters}
\label{1.2}
\begin{eqnarray}
\bar{g}_{\mu\nu} &=& e^{2\varphi} \, g_{\mu\nu}
\label{1.2a} \\
\eta &=& e^{-2\varphi}
\label{1.2b}
\end{eqnarray}
\end{mathletters}%
transforms (\ref{1.1}) into a much simpler expression \cite{3}.
\be
I_G = \frac{1}{4\pi G} \int d^2 x ~ \sqrt{-{g}} \,
\left( \eta R - \lambda \right)
\label{1.3}
\ee
Moreover, since (\ref{1.2a}) describes a conformal redefinition of the metric
and since the matter fields are coupled conformally, the form of the matter
action
does not change with the redefinition (\ref{1.2})
except that
$g_{\mu\nu}$ replaces $\bar{g}_{\mu\nu}$.
But the dynamics implied by (\ref{1.3}) leads to vanishing $R$,
so there is no trace anomaly and no black hole effects,
at least semi-classically.

If one concludes that the conformal trace anomaly interferes with field
redefinition as in (\ref{1.2}),
invalidating the equivalence theorem, so ``that there is not a
unique quantization of dilaton gravity'' \cite{7}, the theory loses all
predictive power, even as the formalism loses descriptive ability.  But it may
be that the above observations on the semi-classical theory are inconclusive.
In this context, one should take note of the published claim that even in the
original formulation (\ref{1.1}) there is no trace anomaly, because any anomaly
can be compensated by a shift in the dilaton field \cite{8}.  [The freedom
of shifting the dilaton field
is especially evident in (\ref{1.3}),
where it is recognized that translating
$\eta$ by a constant changes the action only by the topological term
$\propto  \int d^2 x \, \sqrt{-g} R$.]
Moreover, in a recent calculation the ``black hole'' mass vanishes \cite{9}.
[In Ref.~\cite{9} mass/energy is given a gauge theoretical definition,
and agrees with the ADM value.]

We feel that the gauge principle provides unambiguous direction on how to
proceed through this maze, since gauge invariance resolves quantum field
theoretic ambiguities.  As we shall see, the quantum states that we construct
support the claim that there is no gravity-matter interaction.

In Section \ref{sec:2} we review the model (\ref{1.1}), (\ref{1.2}) in its
gauge theoretical formulation
and describe classical solutions.  The manner in which geometry of space-time
and the trajectory of a particle are encoded in a gauge theory
is noteworthy for its subtlety.
Section \ref{sec:3} is devoted to the formal quantum gauge theory.
Section \ref{sec:4} contains a discussion of the pure gravity
quantum states;
particle states are constructed in Section \ref{sec:5}.
Concluding remarks comprise the last Section VI. 

\section{Gauge~Theory~for~Lineal~Gravity~and~its~classical~solution}
\label{sec:2}

The model that we consider is based on the 4-parameter extended Poincar\'e
group, in (1+1) dimensions, whose Lie algebra is
\begin{eqnarray}
{}~[P_a,\,P_b] &=& \epsilon_{ab} \, I \nonumber\\
{}~[P_a,\,J]   &=& \epsilon_a^{~b} \, P_b \label{2.1}\\
{}~[P_a,\,I]   &=& [J,\,I] = 0 \nonumber
\end{eqnarray}
The central element $I$ modifies the conventional
algebra of translation generators
$P_a$, while the (Lorentz) rotation generator $J$ satisfies conventional
commutators.  Indices $(a,b)=(0,1)$
label a (1+1)-dimensional Minkowski tangent space, with
metric tensor $h_{ab} = {\rm diag~} (1,-1)$, which is used to raise and
lower tangent-space indices.
The anti-symmetric symbol $\epsilon^{ab}$ is normalized by
$\epsilon^{0 1} = 1$.
Although the group is not semi-simple, there
exists an invariant, non-singular bilinear form $P_a P^a - 2 I J$, which
defines a metric tensor on the four-dimensional space of the adjoint
representation.
This metric tensor is used to move indices, so that a four-component
contravariant
vector $V^A$ [$A=(a,2,3)=(0,1,2,3)$]
(transforming with the adjoint representation)
is related to a covariant
vector $V_A$
(transforming with the coadjoint representation)
by $V_a = h_{ab} V^b$, $V_2$ = $-V^3$, $V_3 = - V^2$.
Thus an invariant inner product is defined by
\be
\langle \, W , V \, \rangle
\equiv W_A V^A
= W_a V^a - W_2 V_3 - W_3 V_2
= W^a V_a - W^3 V^2 - W^2 V^3
\label{2.2}
\ee

The gauge theory involves a gauge connection, which is an element of the Lie
algebra,
\be
A_\mu = e_\mu^{\,a} P_a + \omega_\mu J + a_\mu I
\label{2.3}
\ee
and into which are collected
the {\it Zweibein\/} $e_\mu^{\,a}$, the spin-connection
$\omega_\mu$ and a fourth potential $a_\mu$ associated
with the center $I$.  In the usual way, one constructs from (\ref{2.1}) and
(\ref{2.3}) the gauge curvature.
\begin{mathletters}
\label{2.4}
\begin{eqnarray}
F_{\mu\nu} &=& \partial_\mu A_\nu  - \partial_\nu A_\mu + [A_\mu,\,A_\nu]
\label{2.4a} \\
&=& F_{\mu\nu}^{\,a} P_a + F_{\mu\nu}^{\,2} J + F_{\mu\nu}^{\,3} I
\nonumber \\
{\textstyle \frac{1}{2}}
\epsilon^{\mu\nu} F_{\mu\nu} &=&
\epsilon^{\mu\nu}
\Biggl\{
\left( D_\mu e_\nu \right)^a P_a
+
\partial_\mu \omega_\nu J
+
\left( \partial_\mu a_\nu +
{\textstyle \frac{1}{2}} e_\mu^{\,a}
\epsilon_{ab}
e_\nu^{\,b}
\right)
I
\Biggr\}
\label{2.4b}
\end{eqnarray}
\end{mathletters}%
\begin{eqnarray}
\left( D_\mu e_\nu \right)^a
&\equiv&
\partial_\mu e_\nu^{\,a}
+ \epsilon^a_{~b} \omega_\mu e_\nu^{\,b}
\label{2.5}
\end{eqnarray}

These quantities transform with the adjoint representation, whose properties
may be determined from the structure constants of the Lie algebra
(\ref{2.1}).  Given a group element U, then
\begin{eqnarray}
A_\mu \rightarrow A_\mu^{\,U} &=&
 U^{-1} A_\mu \, U + U^{-1} \, \partial_\mu \, U
\label{2.6}\\
F_{\mu\nu} \rightarrow F_{\mu\nu}^{\,U} &=&
 U^{-1} F_{\mu\nu} \, U ~~.
\label{2.7}
\end{eqnarray}
When $U$ is parameterized as
\be
U = e^{\theta^a P_a} e^{\alpha J} e^{\beta I}
\label{2.8}
\ee
with local parameters $(\theta^a, \, \alpha, \, \beta)$,
the transformation (\ref{2.6}) in component form reads
\begin{mathletters}
\label{2.9}
\begin{eqnarray}
e_\mu^{\,a} &\rightarrow&
\left( e^U \right)_\mu^a =
\left( {\Lambda}^{-1} \right)^a_{~b}
\left( e_\mu^{\,b} + \epsilon^b_{~c} \theta^c \omega_\mu +
\partial_\mu \theta^b \right)
\label{2.9a}\\
\omega_\mu &\rightarrow&
\left( \omega^U \right)_\mu =
\omega_\mu + \partial_\mu \alpha
\label{2.9b}\\
a_\mu &\rightarrow&
\left( a^U \right)_\mu =
a_\mu - \theta^a \epsilon_{ab} e_\mu^{\,b} -
{\textstyle \frac{1}{2}} \theta^a \theta_a \omega_\mu + \partial_\mu \beta +
{\textstyle \frac{1}{2}} \partial_\mu \theta^a \epsilon_{ab} \theta^b
\label{2.9c}
\end{eqnarray}
\end{mathletters}
where ${\Lambda}^a_{~b}$ is the Lorentz transformation matrix.
\begin{eqnarray}
{\Lambda}^a_{~b}
&=& \delta^a_{~b} \cosh \alpha + \epsilon^a_{~b} \sinh \alpha
\label{2.10}
\end{eqnarray}

To construct an invariant Lagrange density and action, we introduce a quartet
of Lagrange multiplier fields
\be
\eta_A = \left( \eta_a,\,\eta_2,\,\eta_3 \right)
       = \left( \eta_a,\,-\eta^3,\,-\eta^2 \right)
\label{2.11}
\ee
transforming in the coadjoint representation.
\begin{mathletters}
\label{2.12}
\begin{eqnarray}
\eta_a &\rightarrow&
\left( \eta^U \right)_a =
\left( \eta_b - \eta_3 \, \epsilon_{bc}
\, \theta^c \right)
{\Lambda}^b_{~a}
\label{2.12a}\\
\eta_2 &\rightarrow&
\left( \eta^U \right)_2 =
\eta_2 - \eta_a \, \epsilon^a_{~b} \, \theta^b -
{\textstyle \frac{1}{2}} \eta_3 \, \theta^a \theta_a
\label{2.12b}\\
\eta_3 &\rightarrow&
\left( \eta^U \right)_3 =
\eta_3
\label{2.12c}
\end{eqnarray}
\end{mathletters}%
We then form an invariant by contracting $\eta_A$ with
$\epsilon^{\mu\nu} F_{\mu\nu}^{\,A}$, and take for
the action
\begin{eqnarray}
I_g &=& \frac{1}{4\pi G} \int d^2 x ~
{\textstyle \frac{1}{2}} \epsilon^{\mu\nu}
\left( \eta_a \, F_{\mu\nu}^{\,a} + \eta_2 \, F_{\mu\nu}^{\,2} +
\eta_3 F_{\mu\nu}^{\,3} \right) \nonumber\\
&=&
\frac{1}{4\pi G} \int d^2 x ~
\epsilon^{\mu\nu}
\left( \eta_a
\left( D_\mu e_\nu \right)^a
+
\eta_2 \partial_\mu \omega_\nu +
\eta_3 (\partial_\mu a_\nu + {\textstyle \frac{1}{2}} \, e_\mu^{\,a}
\, \epsilon_{ab} \, e_\nu^{\,b} )
\right)
\label{2.13}
\end{eqnarray}

Since the Lagrange
density involves the gauge invariant inner product
$\langle \eta, \, F_{\mu\nu} \rangle$,
the action is manifestly gauge invariant.
One can show that $I_g$ is equivalent to $I_G$ and $\bar{I}_G$
\cite{3,4}.

A gauge invariant point particle action requires introducing an additional
variable, the ``Poincar\'e coordinate'' $q^a$.  A first order action for a
particle reads \cite{4,10}
\begin{eqnarray}
I_{p} &=& \int d\tau \, \Biggl\{ p_a
\left( D_\tau q \right)^a -
{\textstyle\frac{1}{2}} N (p^a p_a + m^2)
\Biggr\}
\label{2.14}
\end{eqnarray}
\vspace*{-0.35in}
\be
\left( D_\tau q \right)^a \equiv
\skew{-2}\dot{q}^a + \epsilon^a_{~b}
\left( q^b \omega_\mu - e_\mu^{\,b} \right)
\skew{-2}\dot{X}^\mu
\label{2.15}
\ee
The particle dynamical variables are $p_a$, $q^a$ and $X^\mu$, each a function
of $\tau$, which is an affine parameter
--- (\ref{2.14}) is $\tau$-reparametrization-invariant ---
and the over-dot denotes $\tau$-differentiation.
The gravitational variables $\omega_\mu$,
$a_\mu$ and $e_\mu^{\,a}$ in (\ref{2.14}), (\ref{2.15})
are evaluated on the particle trajectory
$X^\mu(\tau)$.
The mass-shell constraint
is enforced by the Lagrange multiplier $N(\tau)$.
The gauge theoretical formalism also accomodates in a very natural manner
various non-minimal gravity-matter interactions mediated by a
velocity dependent interaction with the potentials
$(e_\mu^a, \omega_\mu, a_\mu)$ \cite{4}.  But we do not consider these
elaborations here (see however the discussion in Section \ref{sec:6} and  in
the Appendix).

When the transformation law
for the gravitational variables (\ref{2.9}), (\ref{2.12})
is supplemented
with one for $q^a$ and $p_a$,
\begin{eqnarray}
q^a &\rightarrow&
\left( q^U \right)^a =
({\Lambda}^{-1})^a_{~b}
\, (q^b + \epsilon^b_{~c} \theta^c)
\label{2.16}\\
p_a &\rightarrow&
\left( p^U \right)_a =
p_b \, {\Lambda}^b_{~a}
\label{2.17}
\end{eqnarray}
where the local gauge parameters $(\theta^a,\,\alpha)$ are evaluated on the
particle trajectory $X^\mu(\tau)$,
one finds that the Lagrangian in (\ref{2.14})
is gauge invariant.

[The transformation law (\ref{2.16}) indicates
that $q^a$ comprise the
first two components of a contravariant 4-vector $q^A$, transforming in the
adjoint representation
[{\it i.e.} like (\ref{2.9}) without the derivative terms]
with
$q^2 = -q_3 = 1$
and
$q^3 = -q_2 = {\textstyle\frac{1}{2}} q^a q_a$,
so that $q^A$'s squared ``length''
vanishes, $\langle q,\,q\rangle = 0$.
(The third component of any contravariant vector, equivalently the
fourth component of a covariant vector, is itself always
gauge invariant.)  Similarly, from
(\ref{2.17}) we conclude that $p_a$ comprise the first two components of a
covariant 4-vector $p_A$
transforming in the coadjoint representation
[{\it i.e.\/} like (\ref{2.12})], with vanishing fourth component
$p_3 = p^2 = 0$,
so the squared ``length'' of $p_A$,
$\langle p,\,p\rangle$, is given by $p_a p^a$ and is
constrained by $N$ to be $-m^2$.  A manifestly covariant
formalism and many more details about the extended Poincar\'e group, its
properties and representations are given in Ref.~\cite{4}.]

It is important to notice from (\ref{2.12}) and (\ref{2.16})
that a gauge transformation may be used to set
$\eta_a$ to zero $[\theta^a(x) = \epsilon^{ab} \eta_b(x) / \eta_3(x)]$ or
$q^a$ to zero $[\theta^a \left( X(\tau) \right) = - \epsilon^a_{~b} q^b (\tau)
]$.  In particular, in the gauge $q^a = 0$, the matter action (\ref{2.14})
reduces to the conventional matter-gravity action.
(To recognize this, one should also replace $p_a$ by
$p_b \epsilon^b_{~a}$.)  Thus, we appreciate that the Poincar\'e coordinate is
analogous to the Higgs field in conventional gauge theoretic symmetry
breaking: its presence insures gauge invariance, while a special gauge --- the
unitary gauge (analogous to $q^a=0$) --- exposes physical content.

Equations of motion that follow upon varying the Lagrange multiplier multiplet
$\eta_A = (\eta_a, \eta_2, \eta_3)$ in $I_g$ require vanishing $F_{\mu\nu}$.
\be
F_{\mu\nu} = 0
\label{2.19}
\ee
Varying the gravitational variables
$A_\mu^A = (e_\mu^{a}, \, \omega_\mu, \, a_\mu)$
in $I_g+I_p$ leads to an equation for the Lagrange multiplier multiplet
\be
\partial_\mu \eta + [A_\mu,\,\eta] = 4\pi G \, \epsilon_{\mu\nu} \, J^\nu
\label{2.20}
\ee
where
\be
\eta = \eta^{\,a} P_a - \eta_3 J - \eta_2 I
\label{2.21}
\ee
and the matter current $J^\mu$ is given by
\begin{mathletters}
\label{2.22}
\begin{eqnarray}
J^\mu &=& \int d\tau ~ j \,  \skew{-2}\dot{X}^\mu(\tau) \, \delta^2
\left( x-X(\tau) \right)
\label{2.22a} \\
j &\equiv& j^a P_a + j^2  J + j^3 I \nonumber\\
&=&
-  \epsilon^{ab} p_b \, P_a
- q^a \epsilon_a^{~b} p_b \, I ~~ (j^2 = 0 = j_3)
\label{2.22b}
\end{eqnarray}
\end{mathletters}%
Varying $p_a$ in $I_p$ gives
\be
\left( D_\tau q \right)^a = N p^a
\label{2.23}
\ee
with $p_a$ satisfying the constraint
\be
p_a p^a = -m^2
\label{2.24}
\ee
Varying $q_a$ in $I_p$ leaves, with the help of (\ref{2.23})
\be
\skew{-2}\dot{p}_a =
- \epsilon_a^{~b} \, p_b \, \omega_\mu \, \dot{X}^\mu
\label{2.25}
\ee
Finally, the variation with respect to $X^\mu$ does not produce
an equation independent of the above.

A classical solution to the system is gotten by setting
\be
A_\mu = 0
\label{2.26}
\ee
in order to satisfy (\ref{2.19}),
and the general solution is a gauge transformation of (\ref{2.26}).
With vanishing $A_\mu$,
(\ref{2.25}) becomes $\skew{-2}\dot{p}_a = 0$
and is solved by a constant,
which we choose to write as $\hat{p}_b \epsilon^{b}_{~a}$,
so that $\hat{p}_a$ is timelike when it is
normalized by (\ref{2.24}).
\begin{mathletters}
\begin{eqnarray}
\label{2.27}
p_a &=& \hat{p}_b \epsilon^{b}_{~a} \label{2.27a}\\
\hat{p}_a \hat{p}^a &=& m^2 \label{2.27b}
\end{eqnarray}
\end{mathletters}%
Eq.~(\ref{2.23}) reduces to $\skew{-2}\dot{q}^a = N p^a$
and is solved, using (\ref{2.27}), by
\be
q^a =
\hat{p}_b \epsilon^{ba}
\int\limits^\tau d\tau' ~ N(\tau')
+ \hat{q}^a
\label{2.28}
\ee
where $\hat{q}^a$ are integration constants.
Finally the equations for
$\eta$ are solved after choosing the parameterization
\be
X^0(\tau) = \tau ~~,
\label{2.29}
\ee
by
\begin{eqnarray}
\eta &=& 2\pi G \, \epsilon(\sigma-X(t)) j + \hat{\eta}
\label{2.30}
\end{eqnarray}
since $j$
given in (\ref{2.22b})
is a constant by virtue of (\ref{2.27}) and (\ref{2.28}).
Here once again the $\hat{\eta}$ are integration constants
and $t=x^0, ~ \sigma = x^1, ~ X=X^1$.
Note that $\eta_3$ is gauge invariant and so is the squared ``length'',
$\langle \eta - \hat{\eta}, \, \eta - \hat{\eta} \rangle =
(2 \pi G m)^2$.

The solution as it stands does not define a geometry, because $A_\mu$ and
therefore $e_\mu^{a}$ vanish.  Also the particle trajectory $X(t)$ is
unspecified.  Finally we observe that although a parameterization $\tau$ for
the particle trajectory has been fixed in (\ref{2.29}), $N(\tau)$ remains
undetermined in (\ref{2.28}).  Thus we must answer the question of how
physical information is coded in the above solution.  The answer is subtle.

The physics is found in a new gauge
$A_\mu = U^{-1} \, \partial_\mu \, U$,
where $e_\mu^{a}$ is nonsingular.  At the
same time $q^a$ must be eliminated, and this will determine the orbit $X(t)$.
In other words, the physical content is
exposed in the unitary gauge where $q^a$ vanishes.

It suffices to consider for simplicity gauge transformations in the
translation direction $U = e^{\theta^a P_a}$.  Thus the geometrical
gravitational variables now become from (\ref{2.9}) and (\ref{2.26})
\begin{mathletters}
\label{2.31}
\begin{eqnarray}
e_\mu^{\,a}(x) &=& \partial_\mu \, \theta^a (x)
\label{2.31a}\\
\omega_\mu (x) &=& 0
\label{2.31b}\\
a_\mu(x) &=&
{\textstyle\frac{1}{2}}
\partial_\mu \, \theta^a (x) \, \epsilon_{ab} \, \theta^b (x)
\label{2.31c}
\end{eqnarray}
\end{mathletters}%
According to (\ref{2.17}) the momentum retains its form (\ref{2.27}),
\be
p_a (\tau) = \hat{p}_b \epsilon^{b}_{~a}
\label{2.32}
\ee
while the Poincar\'e coordinate becomes, from (\ref{2.16}) and (\ref{2.28}),
\be
q^a(\tau) =
\hat{p}_b \epsilon^{ba}
\int^\tau d\tau' ~ N(\tau')
+ \hat{q}^a +
\epsilon^a_{~b} \, \theta^b \left( X(\tau) \right) ~~.
\label{2.33}
\ee
Lastly the Lagrange multiplier multiplet $\eta$ reads according to
(\ref{2.12}) and (\ref{2.30})
\begin{mathletters}
\label{2.34}
\begin{eqnarray}
\eta_a(x) &=&
\hat{p}_a
\, 2\pi G \epsilon \left( \sigma - X(t) \right)
+ \left(
\hat{\eta}_a - \hat{\eta_3} \, \epsilon_{ab} \, \theta^{b}(x)
\right)
\label{2.34a}\\
\eta_2 (x) &=&
- \hat{p}_a
\left( \hat{q}^a + \epsilon^a_{~b} \theta^b (x) \right)
\, 2\pi G \epsilon \left( \sigma - X(t) \right) \nonumber\\
&& \qquad
+ \left( \hat{\eta}_2 - \hat{\eta}_a \epsilon^a_{~b} \, \theta^b(x) -
{\textstyle\frac{1}{2}} \hat{\eta}_3 \,
\theta^a (x) \theta_a (x)
\right)
\label{2.34b}\\
\eta_3(x) &=& \hat{\eta}_3
\label{2.34c}
\end{eqnarray}
\end{mathletters}%

While we require that $e^a_\mu = \partial_\mu \theta^a$
be non-singular, there still remains great freedom in fixing its form,
{\it i.e.\/} of selecting $\theta^a$.  A natural choice is $e^a_{\mu} =
\delta^a_{\mu}$, reflecting the fact that $R=0$ and the space-time is flat.
(Of course {\it any\/} form for $\theta^a$ gives a {\it Zweibein\/} that
describes flat space-time.)
Hence we take
\be
\theta^a(x) = x^a
\label{2.35}
\ee
and therefore
\begin{mathletters}
\label{2.36}
\begin{eqnarray}
e_\mu^a &=& \delta_\mu^a \label{2.36a}\\
\omega_\mu &=& 0 \label{2.36b}\\
a_\mu &=& {\textstyle\frac{1}{2}} \, \epsilon_{\mu\nu} \, x^\nu
\label{2.36c}
\end{eqnarray}
\end{mathletters}%
Note that $a_\mu$ is like an electromagnetic vector potential for a constant
field $\epsilon_{\nu\mu}$, which, as is well known, produces a central
extension in the algebra of translations.
(Another choice, popular in the ``black hole'' literature is
$e^a_\mu (t,\sigma) = e^{\lambda\sigma} \delta^a_{\,\mu}$.
To achieve this, it is necessary to perform a local Lorentz gauge
transformation as well as a local translation.)

Once $\theta^a(x)$ is chosen as in (\ref{2.35}), the form of the orbit
$X^\mu(\tau)$ becomes fixed by the requirement that the Poincar\'e coordinate
$q^a(\tau)$ vanishes, {\it i.e.\/} in the ``unitary'', physical gauge.
{}From (\ref{2.33}) and (\ref{2.35}) it follows that
\begin{mathletters}
\label{2.37}
\begin{eqnarray}
\theta^a \left( X(\tau) \right) &=&
\hat{p}^{a} \int^\tau d\tau' \, N(\tau') \, - \epsilon^a_{~b} \hat{q}^b
\label{2.37a}\\
X^a(\tau) &=&
\hat{p}^a \int^\tau d\tau' \, N(\tau') + \hat{X}^a
\label{2.37b}
\end{eqnarray}
\end{mathletters}%
where we have renamed the constant $-\epsilon^a_{~b} \, \hat{q}^b$
as $\hat{X}^a$.

The form of the Lagrange multiplier multiplet is gotten by substituting into
(\ref{2.34}) the $\theta^a(x)$ of (\ref{2.35}) and the
$X^a(\tau)$ of (\ref{2.37b}).
Finally, our choice of parameterization in (\ref{2.29})
and (\ref{2.37b})
fixes $N(\tau)$ to be constant,
\be
N(\tau) = \frac{1}{\hat{p}^0}
\label{2.38}
\ee
so that
\be
X(t) = \pm vt + \hat{X}
\label{2.39new}
\ee
where $v=|{\hat{p}^1/\hat{p}^0}| \leq 1$
and we see that the particle is free.

Thus all aspects of the problem now attain an explicit analytic
and geometric description.
Notice that by virtue of (\ref{2.37a}), where the condition is stated that the
Poincar\'e coordinate vanishes {\bf after} the gauge transformation, the
Poincar\'e coordinate {\bf before} the gauge transformation
(\ref{2.28})
has the same form as
the particle path
(apart from  an $\epsilon$-twist).

The two-particle problem does not provide any new structure.  Upon introducing
an action like (\ref{2.14}) for each particle, we find that there is no
interaction between the particles.
We shall see that in the quantum theory the same physics holds.

\section{QUANTIZATION}
\label{sec:3}

We quantize $I_g + I_p$ using symplectic methods appropriate to
first-order Lagrangians \cite{11} and we solve the constraints as in vector
gauge theories.
In the matter action, we choose the parameterization $X^0(\tau) = \tau$,
so that
there is a common time $t \equiv x^0$ for both the gravity and matter Lagrange
densities, which may be taken as
\begin{eqnarray}
{\cal L} &=& \frac{1}{4\pi G}
\left\{
\eta_a \dot{e}_1^{\,a} + \eta_2 \dot{\omega}_1 + \eta_3 \dot{a}_1
\right\}
+ e_0^{\,a} G_a + \omega_0 G_2 + a_0 G_3
\label{3.1}\\
&&  \quad + \, \Biggl\{
p_a \, \dot{q}^a
+
p_a \, \epsilon^a_{~b}
\left( q^b \omega_1 - e_1^{b} \right) \dot{X}
- \half N (p^a p_a + m^2)
\Biggr\} ~
\delta(\sigma-X)
\nonumber
\end{eqnarray}
where the Gauss constraints $G_A$ read
\begin{mathletters}
\label{3.2}
\begin{eqnarray}
G_a &=& \fpg  \left( \eta'_a + \epsilon_a^{~b} \eta_b \omega_1 + \eta_3
\epsilon_{ab} e_1^{\,b} \right)
+\epsilon_a^{~b} p_b \, \delta (\sigma-X)
\label{3.2a}\\
G_2 = -G^3 &=& \fpg \left( \eta'_2 + \eta_a \epsilon^a_{~b} e_1^{\,b} \right)
- q^a \epsilon_a^{~b} p_b \, \delta(\sigma - X)
\label{3.2b}\\
G_3 = -G^2 &=& \fpg \eta'_3
\label{3.2c}
\end{eqnarray}
\end{mathletters}%
We remind that the fields are functions of $t$ and $x^1 \equiv \sigma$.
The particle variables $p_a$, $q^a$ and $X \equiv X^1$ are functions only of
$t$.  Dot/dash denote respectively differentiation with respect to $t/\sigma$.

{}From (\ref{3.1}) we see that the field ``coordinates'' are
$(e_1^{\,a}, \omega_1, a_1)$,
while their conjugate ``momenta'' are, respectively
$\frac{1}{4\pi G} (\eta_a, \eta_2, \eta_3)$.
Also $p_a$ is
conjugate to the Poincar\'e coordinate $q^a$.
So that $X$ possesses a conjugate momentum, we call
$\Pi$ the coefficient of $\dot{X}$ in (\ref{3.1}),
and enforce that definition with another Lagrange multiplier $u$.
Thus the Lagrange density that we quantize is
\begin{eqnarray}
{\cal L} &=& \fpg
\left(
\eta_a \dot{e}_1^{\,a} + \eta_2 \dot{\omega}_1 + \eta_3 \dot{a}_1
\right)
+
\left(
p_a \dot{q}^a + \Pi \dot{X}
\right)
\, \delta(\sigma - X)
+ e_0^{\,a} G_a + \omega_0 G_2 + a_0 G_3
\nonumber\\
&& \quad
- \, \left\{
\vphantom{\frac{1}{1}}
\half N
\left(
p^a p_a + m^2
\right)
+ u
\left(
\Pi - p_a \epsilon^a_{~b}
(q^b \omega_1 - e_1^{\,b})
\right)
\right\}
\, \delta(\sigma-X)
\label{3.4}
\end{eqnarray}

The algebra of constraints reflects the algebraic underpinnings of the theory.
The four Gauss law generators reproduce the Lie algebra (\ref{2.1}).
The non-vanishing commutators are as expected
\begin{mathletters}
\label{3.5}
\begin{eqnarray}
{}~[ G_a (\sigma), G_b (\sigma') ]
&=& i \, \epsilon_{ab} \, G_3 (\sigma) \,  \delta(\sigma-\sigma')
\label{3.5a} \\
{}~[ G_a (\sigma), G_2 (\sigma') ]
&=& i \, \epsilon_a^{~b} \, G_b (\sigma) \, \delta(\sigma-\sigma')
\label{3.5b}
\end{eqnarray}
\end{mathletters}%
(In fact the above commutators, valid for any coupling constant $4\pi G$, hold
separately for the gravity part and for
the matter part of $G_A$.)  Moreover the mass shell constraint (enforced by
$N$) and the momentum constraint (enforced by $u$) commute with $G_A$
and with each other.  Thus all the constraints are first-class and can be
imposed as conditions on physical quantum states.  This we now proceed to do,
to begin with
in the next Section
just for the gravity portion and then, in the following Section,
for the combined gravity-particle system.

\section{GRAVITATIONAL STATES}
\label{sec:4}

In this Section, we delete the matter (particle) variables and discuss the
quantum states of pure gravity \cite{5,6}.  {}From (\ref{3.4}) it is seen that
the Hamiltonian
density consists solely of the Gauss constraints $G_A=(G_a,\,G_2,\,G_3)$
enforced by $A_0^{\,A} = (e_0^{\,a}, \omega_0, a_0)$.
Since the algebra (\ref{3.5}) shows the constraints to be first-class, they
may be imposed on states, and the quantum theory has no further structure.
Before imposing the Gauss law constraints, let us first discuss in greater
detail how gauge transformations act in the quantum theory.

Examining the explicit expressions for the $G_A$, we recognize that they
generate by commutation the relevant gauge transformations on the dynamical
variables, {\it i.e.\/} the infinitesimal forms of (\ref{2.9}) and
(\ref{2.12}).  However, we further note that whereas the full generators are
needed to implement the gauge transformation on the ``coordinates''
$A_1^A = (e_1^{a}, \omega_1, a_1)$,
the derivative parts of generators
$\propto (\eta'_a, \eta'_2, \eta'_3)$
commute with the ``momenta''
$\eta_A \propto (\eta_a, \eta_2, \eta_3)$
and are not needed for effecting the gauge transformation on the ``momenta''.
(This of course merely reflects the circumstance that the ``coordinates'' are
connections, which experience inhomogenous gauge transformations, while the
``momenta'' transform covariantly.)

A consequence of this difference emerges when we consider,
before enforcing the Gauss law, quantum states in
the Schr\"odinger representation as functionals either of the ``coordinates''
or the ``momenta''.
Let us act on such functionals
with the unitary operator ${\cal U}$ that implements a finite gauge
transformation $U$.
\be
{\cal U} = e^{i \int d\sigma \, \theta^a G_a}
{}~ e^{i \int d\sigma \, \alpha G_2 }
{}~ e^{i \int d\sigma \, \beta G_3}
\label{4.1}
\ee
Acting on functionals of ``coordinates''
({\it i.e.\/} connections $A_1^{\,A}$),
${\cal U}$ gauge transforms the argument of the functional.
However, when ${\cal U}$
acts on functionals of ``momenta''
({\it i.e.\/} Lagrange multipliers $\eta_A$), in addition to a gauge
transformation on the argument of the functional, there arises a
multiplicative phase.  This can also be seen from the Fourier transform
relation between functionals of ``coordinates''
$\Phi (A_1)$ and functionals of ``momenta'' $\Psi(\eta)$ \cite{12}.
\begin{eqnarray}
\Psi(\eta) &=& \int {\cal D} A_1 \, e^{-\frac{i}{4\pi G} \int d\sigma \,
\langle \eta, A_1 \rangle} \, \Phi (A_1)
\label{4.2} \\
{\cal U}^{-1} \, \Psi(\eta) &=& \int {\cal D} A_1
e^{-\frac{i}{4\pi G} \int d\sigma \,
\langle \eta, A_1 \rangle} \, {\cal U}^{-1} \Phi (A_1) \nonumber\\
&=&
e^{\frac{i}{4\pi G} \int
\langle \eta, dU \, U^{-1} \rangle}
\, \int {\cal D} A_1 \,
e^{-\frac{i}{4\pi G} \int d\sigma \,
\langle \eta, U A_1 U^{-1} \rangle} \, \Phi(A_1) \nonumber\\
&=&
e^{\frac{i}{4\pi G} \int
\langle \eta, dU \, U^{-1} \rangle}
\,
\Psi(\eta^U)
\label{4.3}
\end{eqnarray}

Of course the Gauss law demands that physical states be annihilated by the
generators
$G_A$
and left invariant by ${\cal U}$.  Thus, states in the ``coordinate''
representation are gauge invariant, while those in the ``momentum''
representation are gauge invariant up to a phase, {\it i.e.\/} they
satisfy, according to (\ref{4.3}),
\begin{equation}
\Psi
\left(
\eta^U
\right) =
e^{-\frac{i}{4\pi G} \int  \langle \eta, \, dU \, U^{-1} \rangle} \,
\Psi(\eta)
\label{4.4}
\end{equation}

It turns out to be more convenient to work in the ``momentum''
representation,
so we seek  functionals that obey (\ref{4.4}),
with $\eta^U$ given in (\ref{2.12}).   Such functionals
are readily constructed by satisfying the
infinitesimal version of (\ref{4.4}), {\it i.e.\/}
by solving
the constraint that Gauss
generators (\ref{3.2}) annihilate physical states.
\begin{mathletters}
\label{4.5}
\begin{eqnarray}
\left(
\eta'_a(\sigma) + i \, 4\pi G \, \epsilon_a^{~b} \eta_b (\sigma)
\, \frac{\delta}{\delta \eta_2(\sigma)}
+ i \, 4\pi G \, \eta_3(\sigma) \, \epsilon_{ab} \,
\frac{\delta}{\delta \eta_b(\sigma)} \right)
\Psi(\eta) &=& 0
\label{4.5a} \\
\left( \eta'_2 (\sigma) + i \, 4\pi G \, \eta_a(\sigma) \,
\epsilon^a_{~b} \frac{\delta}{\delta \eta_b (\sigma)} \right)
\Psi(\eta) &=& 0
\label{4.5b} \\
\eta'_3(\sigma) \,
\Psi(\eta) &=& 0
\label{4.5c}
\end{eqnarray}
\end{mathletters}%
The solution to these equations is
\be
\Psi(\eta) =
\delta(\eta'_3) \,
\delta \left( \left[ \eta^a \eta_a - 2 \eta_2 \eta_3 \right]' \right)
\, e^{i\Omega} \, \psi
\label{4.6}
\ee
where $\psi$ depends in an arbitrary fashion on the constant parts of the
invariants $\eta^a \eta_a - 2 \eta_2 \eta_3$ and $\eta_3$,
and the phase $\Omega$ is given by
\be
\Omega = \frac{1}{8\pi G} \int \epsilon^{ab} \eta_a \, d\eta_b
\Bigl/ \eta_3
\label{4.6ex}
\ee
The only gauge non-invariant portion of (\ref{4.6}) is its phase, and one
easily
confirms that under the gauge transformation (\ref{2.12}), (\ref{4.4}) is
true.
The phase may be reexpressed by noting that
$\eta_3$ is an invariant, only whose
constant part survives in (\ref{4.6}); call  it  ${\lambda}$.
Thus physical no-particle states are described by states of
the form
\be
\Psi \sim \exp \left[
{\frac{i}{8 \pi G \lambda} \int \, \epsilon^{ab} \eta_a d\eta_b}
\right]  \,
{}~
\psi(M,\lambda)
\label{4.7}
\ee
where $M$ is the constant part of the invariant
$\eta_a \eta^a - 2 \eta_2 \eta_3$.
When reference is made to the geometrical formulation of the model,
{\it e.g.\/} (\ref{1.3}), it is established that $\lambda$ is just the
cosmological constant.  In the {\bf gauge} theory, this is not a parameter,
but a possible value of a dynamical variable \cite{3,4}.
Also $M$ plays the role of the
``black hole'' mass in the classical solution \cite{1,2}; in the quantized
gauge theory it too is a variable.

The phase (\ref{4.6ex}) has the following group theoretical significance.

It is known that the Lie algebra for a group can be obtained
from the canonical 1-form
$\langle K, \, dg \, g^{-1} \rangle$.
Here $K$ is a constant element of the Lie
algebra, $g$ a group element and $\langle \, , \,  \rangle$ defines an
invariant inner product on the Lie algebra.  [For semi-simple groups this
would be the Cartan-Killing metric; otherwise
--- for example in our extended Poincar\'e group ---
we use another metric, as in (\ref{2.2}).]
When the group generators $Q$ are defined to be
$Q = g^{-1} K g $, one finds that their Poisson brackets,
as determined by the above 1-form
and by the symplectic 2-form
$d \langle K, \, dg \, g^{-1} \rangle =
\langle K, \, dg \, g^{-1} dg \, g^{-1} \rangle$,
reproduce the Lie algebra.  The 2-form is the Kirillov-Kostant  symplectic
2-form, and we similarly name the 1-form.
(This 1-form in general is well defined only locally.)

We now show that $\Omega$ is precisely
$\int \langle K, dg \, g^{-1} \rangle$, where
$K$ is any fixed element in the maximal abelian subalgebra spanned by the
generators $J,I$ of the extended Poincar\'e group and
$\eta$ is identified with
\be
\eta = g^{-1} \, K \, g ~~.
\label{4.8}
\ee
We require that under a gauge transformation $U$,
$K$ is invariant while $g$ transforms as
$g \to g^U = gU$, so that $\eta \to \eta^U = U^{-1} \eta U$.
It follows that
\begin{eqnarray}
\langle \, K , \, dg \, g^{-1} \, \rangle
&\to& \langle
\, K , \, dg \, g^{-1} \, + g \, dU \, U^{-1} \, g^{-1}  \,
\rangle
\nonumber\\
&=& \langle \, K , \, dg \, g^{-1} \, \rangle +
\langle \, g^{-1}  \, K \, g , \, dU \, U^{-1} \, \rangle
\nonumber\\
&=& \langle \, K , \, dg \, g^{-1} \, \rangle +
\langle \, \eta , \, dU \, U^{-1} \, \rangle  ~~.
\label{4.9}
\end{eqnarray}
Hence
$\Psi(\eta) \propto e^{\frac{i}{4\pi G} \int \langle K , \,
dg \, g^{-1} \rangle}
\, \psi(M,\lambda)$ transforms as required by (\ref{4.4})
with $K = - \lambda J + \frac{M}{2\lambda} I$ and
$g = \exp (\eta_a \epsilon^{ab} P_b / \lambda)$.
Notice that $g$ is defined in (\ref{4.8}) only up to a left multiplication
by a element $h$ in the maximal abelian subgroup.  By replacing $g$ with $hg$
the phase is shifted by the boundary term
${i \over 4 \pi G} \int d \langle K, \ln h \rangle$,
which may induce topological effects.

Explicit evaluation, when $U$ is given as in (\ref{2.8}), confirms the above,
and $\eta_a$ parameterizes the two-dimensional (co-) adjoint orbit of the
group; indeed the $\eta_a$ are just the Darboux (canonical) coordinates
on the reduced phase space \cite{13}.

\section{WAVE FUNCTIONALS IN THE PRESENCE OF MATTER}
\label{sec:5}

In this Section we extend the results of Section \ref{sec:4}, by including
matter degrees of freedom, to begin with a single point particle.  We remain
with the ``momentum'' representation for the gravity variables, but describe
the particle by position variables, so the state is a functional of $\eta_A$
and a function of $q^a$ and $X$, while $p_a$ and $\Pi$ are realized by
differentiation.  The gauge transformation generators now include a matter
contribution $j_A$, and their exponential
acting on arbitrary functionals again gauge transforms the argument,
and multiplies the wave functional with the same phase as in (\ref{4.3}).
\be
{\cal U}^{-1} \Psi (\eta, q, X) =
e^{\frac{i}{4\pi G} \int \langle \eta, dU \, U^{-1} \rangle} \,
\Psi \left( \eta^U, q^U, X \right)
\label{5.1}
\ee
Thus application of the Gauss law, which requires the left side of
(\ref{5.1})
to be
$\Psi (\eta,\, q,\,X)$, constrains the wave functional to satisfy
\be
\Psi \left( \eta^U, \, q^U, X \right) =
e^{-\frac{i}{4\pi G} \int \langle \eta , \, dU \, U^{-1} \rangle}
\, \Psi(\eta,\, q,\, X)
\label{5.2}
\ee

Once again solving this constraint is best accomplished from its
infinitesimal
version.  We impose the requirement that the Gauss law generators
annihilate the state.  The resultant differential equations are as in
(\ref{4.5}), except the right sides now contain matter contributions.
\begin{mathletters}
\[
\left(
\eta'_a(\sigma) + i 4\pi G \, \epsilon_a^{~b} \, \eta_b(\sigma) \,
{\delta \over \delta\eta_2(\sigma)} + i 4\pi G \, \eta_3(\sigma) \,
\epsilon_{ab} \, {\delta \over \delta \eta_b(\sigma)}
\right) \, \Psi(\eta,q,X)
\hspace*{\fill}
\]
\vspace*{-0.5in}
\begin{equation}
\hspace*{1.75in}
= - 4 \pi G \, \delta(\sigma - X) \, \epsilon_a^{~b}
\frac{1}{i} \, \frac{\partial}{\partial q^b}
\,
\Psi (\eta,q,X)
\label{5.3a}
\end{equation}
\begin{eqnarray}
\left(
\eta'_2 (\sigma) + i 4 \pi G \, \eta_a(\sigma) \,
\epsilon^a_{~b} \, \frac{\delta}{\delta \eta_b(\sigma)}
\right)
\, \Psi(\eta,q,X) &=& 4 \pi G \, \delta(\sigma - X)
\, q^a \epsilon_a^{~b} \,
\frac{\partial}{i \, \partial q^b}
\, \Psi(\eta,q,X)
\label{5.3b} \\
{\vrule width 0pt height 20pt}
\eta'_3 (\sigma) \, \Psi(\eta,q,X)
&=&
0
\label{5.3c}
\end{eqnarray}
\end{mathletters}%
Eqs.~(\ref{5.3a}) and (\ref{5.3c}) are solved by
\begin{eqnarray}
\Psi(\eta,q,X) &=& e^{i\Omega} \, \delta
\Bigl( \eta'_3 \Bigr)
\, \hat{\Psi} (\eta_a \eta^a - 2 \eta_2 \eta_3 ,
\lambda, \rho , X )
\label{5.4}
\end{eqnarray}
where the phase is given as before by (\ref{4.6ex})
and $\rho$ is defined by
\be
\rho^a = q^a + \eta^a(X) / \eta_3 (X)
\label{n5.6}
\ee
By virtue of (\ref{5.3c}) and
the functional $\delta$-function in
(\ref{5.4}), $\eta_3$ is the constant $\lambda$.

Imposing the remaining constraint (\ref{5.3b})
leads to the following equation for ${\Phi}$.
\be
\left\{
\left(
{\eta^a \eta_a - 2 \eta_2 \eta_3}
\right)' (\sigma)
+ 8 \pi G \lambda \,
\delta (\sigma - X)
\,
\rho^a \epsilon_a^{~b}
\frac{\partial}{i \, \partial \rho^b}
\right\}
\hat{\Psi} = 0
\label{n5.7}
\ee
Note that $\rho$ responds only to the Lorentz rotation part of a general gauge
transformation.  Hence (\ref{n5.7}) is a gauge invariant equation, and so also
is (\ref{5.3c}).

Solving (\ref{n5.7}) is accomplished by diagonalizing the operator
$\rho^a \epsilon_a^{~b} \frac{\partial}{i \, \partial \rho^b}$
so that it acquires the [continuous] eigenvalue $\nu$.
\be
\rho^a \epsilon_a^{~b}
\frac{\partial}{i\,\partial \rho^b} \hat{\Psi} = \nu \,
\hat{\Psi}
\label{dc5.12}
\ee
This fixes the ``angular'' $\rho$ dependence
of ${\Psi}$ and then (\ref{n5.7}) is solved by a functional
$\delta$-function that evaluates
${\eta^a \eta_a - 2 \eta_2 \eta_3}$,
leaving still undetermined an arbitrary gauge invariant function of
$\rho_a \rho^a$ and $X$.
\begin{eqnarray}
&& \Psi(\eta,\rho,X) \label{n5.8} \\
&& \hbox{\quad} =
e^{i\Omega} \,
\delta \Bigl( \eta'_3 \Bigr) ~
\delta \left(
\left(
{\eta^a \eta_a - 2 \eta_2 \eta_3}
\right)'
+ 8 \pi G  \lambda \, \nu \, \delta (\sigma - X)
\right)
\,
\left(
\frac{\rho^0 + \rho^1}{\rho^0 - \rho^1}
\right)^{i\nu/2}
\psi_\nu (M, \lambda, \rho^a \rho_a, X)
\nonumber
\end{eqnarray}
where $M$ is the constant part of the invariant $\eta^a\eta_a-2\eta_2\eta_3$.

It now remains to solve the momentum and mass shell constraints.  We consider
first the former --- as will be seen it does not lead to any new structure,
but merely eliminates the $X$-dependence of $\psi$.  That constraint reads
\begin{mathletters}
\begin{equation}
\left(
\Pi + \omega_1 (X) \, q^a \epsilon_a^{~b} p_b + p_a \epsilon^a_{~b} e_1^b
(X)
\right)
\Psi = 0
\label{n5.9a}
\end{equation}
Since $\Psi$ satisfies the translational constraint (\ref{5.3a}),
$e_1^a (X) \Psi$
may be evaluated from that equation, whereupon
(\ref{n5.9a}) becomes
\begin{equation}
\left(
\Pi - \frac{\eta'_a (X)}{\eta_3(X)} p^a
+ \omega_1(X) \rho^a \epsilon_a^{~b} p_b
\right)
\Psi = 0
\label{n5.9b}
\end{equation}
Next moving the phase and the functional $\delta$-functions that are present
in $\Psi$
across the operator in (\ref{n5.9b})
and evaluating
$\rho^a \epsilon_a^{~b} \frac{\partial}{i \, \partial \rho^b}$
on its eigenvalue $\nu$ exposes the non-trivial content of the momentum
constraint, as a differential equation for
$\psi_\nu (M, \lambda, \rho^a \rho_a, \, X)$.
\be
\left(
\frac{1}{i} \, \frac{\partial}{\partial X} -
\frac{\eta'_a(X)}{\lambda} \, \frac{\partial}{i \, \partial \rho^a}
\right)
\,
\left(
\frac{\rho^0 + \rho^1}{\rho^0 - \rho^1}
\right)^{i\nu/2}
\psi_\nu (M, \lambda, \rho^a \rho_a, X) = 0
\label{n5.9c}
\ee
\end{mathletters}%
Since $\rho^a$ depends on $X$ through its definition (\ref{n5.6})
[with $\eta_3(X) = \lambda$], we see that (\ref{n5.9c}) merely states that
$\psi_\nu$ has no explicit $X$ dependence.
Thus the one-particle gravitational state is described by the functional
\begin{eqnarray}
\Psi(\eta,q,X) &=&
\exp\left({\frac{i}{8\pi G \lambda} \int \epsilon^{ab} \eta_a d\eta_b} \right)
\,
\delta \Big( \eta'_3 \Big)
\,
\delta
\left(
\left(
{\eta^a \eta_a - 2 \eta_2 \eta_3}
\right)'
+ 8 \pi G \lambda \, \nu \, \delta(\sigma - X)
\right)
\nonumber \\
&& \hspace{0.6in} \times \,
\left(
\frac{\rho^0 + \rho^1}{\rho^0 - \rho^1}
\right)^{\frac{i}{2}\nu}
\,
\psi_\nu (M, \lambda, \rho^a \rho_a)
\label{n5.10}
\end{eqnarray}
with the gauge-invariant function $\psi_\nu$
to be determined by the mass shell constraint.

This last constraint is enforced by $N$ in Eq.~(\ref{3.4}),
\be
\left( - \frac{\partial}{\partial \rho^a} \frac{\partial}{\partial \rho_a} +
m^2 \right) \Psi = 0
\label{dc5.11}
\ee
and with the diagonalization (\ref{dc5.12})
it implies a second order differential equation for $\psi_\nu$.
\be
\left[ \frac{d^2}{dz^2} + \frac{1}{z} \frac{d}{dz} - \left( \frac{m}{2}
\right)^2 \frac{1}{z} + \left( \frac{\nu}{2} \right)^2 \frac{1}{z^2} \right] \,
\psi_\nu(z) = 0
\label{dc5.13}
\ee
The two solutions are Bessel functions of the second type.
\be
\psi_\nu(M, \lambda, \rho^a \rho_a) \propto \left\{ \begin{array}{l}
              I_{i\nu}(m \sqrt{\rho^a \rho_a}) \\
              K_{i\nu}(m \sqrt{\rho^a \rho_a}) \end{array}
\right.
\label{dc5.14}
\ee
They differ in their asymptotic behavior: for large positive value of $\rho^a
\rho_a$, the function $I_{i\nu}$ diverges exponentially while $K_{i\nu}$ decays
exponentially.
We saw in Section \ref{sec:2} that the classical solution does not specify the
classical path until the physical gauge $q^a = 0$ is chosen.   Alternatively
with $q^a \neq 0$ but in the nonsingular gauge where $e_\mu^a \neq 0$, one may
identify the classical trajectory with $-\epsilon^a_{~b} q^b$.   Since the
quantal wave function depends on $\rho^a = q^a + \eta^a/\lambda$, we may
interpret $\rho^a \rho_a$ as $X^2 - t^2$.  The physical requirement that wave
functions do not diverge at large distance would then disallow the
$I$-solution.
This point, as well as the similarity of
the quantal description to a free particle in 1+1 dimensions,
are detailed in the Appendix.

Let us briefly comment on the case of several matter particles. We add in the
action one interaction term (\ref{2.14}) per particle. The different masses,
trajectories and momenta are labeled by an index: $m_{(n)}$, $X_{(n)}$,
$\Pi_{(n)}$. We also introduce
a Poincar\'e coordinate
$q^a_{(n)}$ (and the
corresponding momentum $p_a^{(n)}$) per particle.
(Alternatively, we can view $q^a$ as a function of space-time, which enters in
this system only through its value on the trajectories, $q^a(t, X_{(n)})
\equiv q^a_{(n)}(t)$. In the case of field theory, the complete function
$q^a(t, \sigma)$ would appear \cite{4}.)
Notice that no specific non-gravitational interaction between the particles
has been considered.

The Gauss laws (\ref{5.3a}) and (\ref{5.3b}) have now on the right side
a sum of
$\delta$-functions peaked on the different trajectories and we get one momentum
constraint~(\ref{n5.9a}) and one mass shell contraint~(\ref{dc5.11}) for each
particle. The physical state for several particles thus ``factorizes'' and is
\begin{eqnarray}
\Psi(\eta,q_{(n)},X_{(n)}) &=&
\exp
\left[ {\frac{i}{8\pi G \lambda} \int \epsilon^{ab} \eta_a d\eta_b} \right]
\delta \Big( \eta'_3 \Big) ~
\delta \left(
( \eta^a \eta_a - 2 \eta_2 \eta_3 )'
+ 8 \pi G \, \lambda \sum_n \, \nu_{(n)} \, \delta(\sigma - X_{(n)})
\right)
\nonumber \\
&& \quad \times \, \prod_n
\,
\left(
\frac{\rho^0_{(n)} + \rho^1_{(n)}}{\rho^0_{(n)} - \rho^1_{(n)}}
\right)^{i\nu_{(n)}/2}
K_{i \nu_{(n)}} (m_{(n)} \sqrt{\rho^a_{(n)} \rho_{a (n)}})
\label{dc5.15} \\[12pt]
\rho^a_{(n)} &=& q^a_{(n)} + \eta^a(X_{(n)}) / \lambda
\nonumber
\end{eqnarray}
which indicates that there is no interaction between the particles.

\section{DISCUSSION}
\label{sec:6}

Our quantization procedure does not give evidence of any gravitational force
between the matter particles moving on a line.
We believe that this conclusion cannot be avoided, as long as gauge invariance
is maintained.  The possibility of ``not $\ldots$ unique quantization of
dilaton gravity'' \cite{7} is eliminated by the gauge principle.

Let us however call attention to a subtle effect, that is not apparent in what
has been done above, but may be relevant in other situations.  The effect
that we wish to discuss is most readily seen in the gauge $\eta_a=0$.
When this gauge is elected,
the wave functional simplifies to $\delta (\eta'_3) \,
\delta \left(\eta'_2 - 4 \pi G \lambda \, \nu \, \delta (\sigma - X) \right)
\left( {q^0 + q^1 \over q^0 - q^1} \right)^{\,\frac{i}{2}\nu}
K_{i \nu} \left( m \sqrt{q^a q_a} \right)$,
the translational gauge freedom generated by $G_a$ is fixed,
but one must also take
into account the non-trivial nature of the
$\left[ G_a(\sigma), \, \eta_b(\sigma') \right] $ bracket, which is
$i \epsilon_{ab} \eta_3 (\sigma) \, \delta (\sigma - \sigma')$.
Since
$\ln \det \left[ G_a, \eta_b \right]$
is effectively
$2 \delta(0) \int d \sigma \, \ln \eta_3 (\sigma)$,
the wave functional possesses a further factor
$e^{- \delta(0) \int d\sigma \, \ln \eta_3 (\sigma)}$.
In our case this factor
is invisible because $\eta_3$ is, according to (\ref{5.3c}),
the constant $\lambda$, and the factor is an irrelevant constant.  Constancy
of $\eta_3$, in the presence of matter, is a consequence of the
absence of a
matter coupling to $a_\mu$; {\it viz.\/} $j_3$ vanishes.
However, as we have remarked already, it is
possible to introduce a non-minimal matter gravity interaction
${\cal B} a_\mu \dot{X}^\mu$, which changes $\eta_3(\sigma)$ to
$\lambda + 2 \pi G {\cal B} \, \epsilon(\sigma - X)$.
The finite part of $\int d\sigma \ln \eta_3 (\sigma)$ may be evaluated
 by first differentiating with respect to $X$,
\[
{d \over dX} \int d \sigma
\, \ln \eta_3 (\sigma) = - \int {d\sigma\over \eta_3(\sigma)} \,
{4 \pi G {\cal B} \, \delta (\sigma - X)}
= - {4 \pi G {\cal B} \over \lambda} ~~,
\]
and we conclude that  the wave functional acquires the singular factor
$\exp  \left( {\delta (0) \over \lambda} \, 4 \pi G {\cal B} X \right)$.
[In fact the same factor emerges when the calculations of
Section~\ref{sec:5} are repeated without choosing the $\eta_a=0$ gauge
but
in the
presence of the non-minimal ${\cal B} a_\mu \dot{X}^\mu$ interaction.
Specifically the singular factor is encountered when solving the momentum
constraint.]
We do not know how to assess this singularity, which, to reiterate, does not
affect the model considered in the body of this paper.

\section*{ACKNOWLEDGEMENT}

We thank Dong-Su Bak and Washington Taylor IV for instructive conversations
about the Kirillov-Kostant theory.

\section*{APPENDIX}

\makeatletter
\def\theequation@prefix{A.}
\makeatother

In this Appendix we present a quantal description for the free motion of a
relativistic particle in (1+1)-dimensional space-time.  Our purpose is to
exhibit in this familiar context formulas identical to those in the body of
the paper derived from ``string-inspired'' gravity.

The Lagrangian is
\begin{equation}
L_{\rm particle} = -\Pi_\mu \dot{X}^\mu - \frac{N}{2}
\left( \Pi^\mu \Pi_\mu - m^2 \right)
\label{A.1}
\end{equation}
It contains the mass-shell constraint and is parameterization invariant.  One
may quantize in a parameterization invariant fashion, imposing the constraint
on covariant wave functions with $\Pi_\mu$ replaced by
$\frac{\partial}{i \partial X^\mu}$.  In this way one is led to the equation
\be
\left( \square + m^2 \right)
\, \psi(X) = 0
\label{A.2}
\ee
Since the Lorentz generator in this theory,
${\cal M} = X^\mu \epsilon_\mu^{~\nu} \Pi_\nu$,
commutes with the mass-shell constraint, it may be additionally
imposed.
\be
 X^\mu \epsilon_\mu^{~\nu}
\, \frac{\partial}{i \partial X^\nu}
\psi(X) = - \nu \, \psi(X)
\label{A.3}
\ee
Clearly (\ref{A.1}) and (\ref{A.2}) are identical with (\ref{dc5.11})
and (\ref{dc5.12}); they possess the
solution (\ref{dc5.14}) with $X^\mu$ identified with
$-\epsilon^\mu{}_\nu \rho^\nu$.

An alternative point of view, within which one may also justify the selection
of the $K$-Bessel solution over the $I$-Bessel solution, is provided by solving
the
constraint first and choosing the parameterization
$X^0(\tau) = \tau$.  We then have $\Pi_0 = \sqrt{\Pi_1^{\,2} + m^2}$ and
\begin{equation}
L_{\rm particle} \to \Pi \dot{X} - \sqrt{\Pi^2 + m^2}
\label{A.4}
\end{equation}
where $X^0=t$, $X^1\equiv X$, $\Pi_1 \equiv \Pi$, with $X$ and $\Pi$ carrying
a $t$-dependence.
We are not interested in energy eigenstates.  Rather we seek to diagonalize
the Lorentz generator, which in the parameterized formalism reads
\begin{equation}
{\cal M} = - t\Pi + X \sqrt{\Pi^2 + m^2}
\label{A.5}
\end{equation}

Solution of the Lorentz eigenvalue problem in $X$-space is difficult owing to
the non-locality of the energy operator.  Therefore we introduce the
momentum-space wave functions $\varphi(t,p)$ and impose the symmetrized
version of (\ref{A.5}) as an eigenvalue condition.
\begin{equation}
\left(
- t p + \frac{i}{2} \frac{\partial}{\partial p} \sqrt{p^2 + m^2}
+ \sqrt{p^2 + m^2}
\frac{i}{2} \frac{\partial}{\partial p}
\right)
\varphi_\nu = -\nu \varphi_\nu
\label{A.6}
\end{equation}
The solution of this first order differential equation is unique,
\begin{equation}
\varphi_\nu (t,p) =  e^{-it \sqrt{p^2 + m^2}}
{}~ \frac{(p + \sqrt{p^2 + m^2})^{i\nu}}{(p^2 + m^2)^{1/4}}
\label{A.7}
\end{equation}
where a normalization constant is fixed by
\begin{equation}
\int \frac{dp}{2\pi} ~
\varphi^{*}_{\nu'} (t,p) \,
\varphi_\nu (t,p) =
\delta(\nu-\nu') ~~.
\label{A.8}
\end{equation}

We wish to compare this with the solution within the
parameterization independent formalism.  To that end, define the transform
\begin{equation}
\psi_\nu(t,X) = \int \frac{dp}{2\pi} \,
\frac{e^{ip X}}{(p^2 + m^2)^{1/4}} \,
{\varphi_\nu(t,p)}
\label{A.9}
\end{equation}
The reason for the additional factor of $(p^2 + m^2)^{-1/4}$
in the measure is understood as follows.   If $\psi_\nu(t,X)$ is to be
identified with the parameterization independent solution, it should satisfy
the Klein-Gordon equation, and indeed the function
$\psi_\nu$
in (\ref{A.9})
does so, since the time-dependence of $\varphi_\nu$ is $e^{-it\sqrt{p^2+m^2}}$.
Klein-Gordon solutions are normalized by
\begin{eqnarray}
\delta(\nu-\nu') &=& \frac{i}{2} \int dX
\left(
\psi^{*}_{\nu'} (t,X) \,
\dot{\psi}_\nu (t,X) -
\dot{\psi}^{*}_{\nu'} (t,X) \,
\psi_\nu(t,X)
\right)
\label{A.10}
\end{eqnarray}
and this is seen to require the measure as in (\ref{A.9})
when $\varphi_\nu$ is normalized by (\ref{A.8}).

Carrying out the integral (\ref{A.9}) gives
\begin{equation}
\psi_\nu(t,X) =
e^{ \mp \nu \pi / 2}
\frac{m^{i\nu}}{\pi} \,
\left( \frac{X + t}{X - t} \right)^{i\nu/2} \,
K_{i\nu} (m \sqrt{X^2 - t^2})
\label{A.11}
\end{equation}
with the upper (lower) sign if $(X - t) > 0$ (resp. $(X - t) < 0$).
Thus we arrive unambiguously at
the solution that is well-behaved in space-like
directions, and in this way motivate
the choice made in the text of discarding the $I$ solution.

Finally we remark that the addition of the non-minimal ${\cal B} a_\mu
\dot{X}^\mu$ interaction to the quantum field
theory results in a wave functional that
coincides, apart from the previously mentioned singular factor, with the wave
function of a particle moving in an external electric field ${\cal B}$, in
flat (1+1) dimensional space-time.




\goodbreak

\end{document}